%% file: smcs.tex
\documentclass[12pt]{article}
\usepackage{graphicx}

\newcommand{\be}{\begin{equation}}
\newcommand{\ee}{\end{equation}}
\newcommand{\bea}{\begin{eqnarray}}
\newcommand{\eea}{\end{eqnarray}}
\newcommand{\beann}{\begin{eqnarray*}}
\newcommand{\eeann}{\end{eqnarray*}}
\newcommand{\nn}{\nonumber}
\newcommand{\Tr}{{\rm Tr}}
\newcommand{\del}{\partial}
\newcommand{\rot}[3]{\left[{#1}\atop{#2}\right]_{#3}}
\newcommand{\Z}{{\bf Z}}
\newcommand{\til}{\widetilde}

\makeatletter
  
  \@addtoreset{equation}{section}
\makeatother

\begin{document}

%
%
\begin{titlepage}

\setcounter{page}{0}
\renewcommand{\thefootnote}{\fnsymbol{footnote}}

\begin{flushright}
YITP-99-23 \\
hep-th/9904118
\end{flushright}

\vspace{15mm}
\begin{center}
{\Large\bf
Moduli Space of Vacua of Supersymmetric Chern-Simons Theories and
Type IIB Branes
}

\vspace{15mm}
{\large
Kazutoshi Ohta\footnote{e-mail address:
kohta@yukawa.kyoto-u.ac.jp}
}\\
\vspace{10mm}
{\em Yukawa Institute for Theoretical Physics, Kyoto University,
Kyoto 606-8502, Japan} \\
\end{center}
\vspace{15mm}
\centerline{{\bf{Abstract}}}
\vspace{5mm}

We study the correspondence between the moduli space of vacua of
three-dimensional supersymmetric Yang-Mills (Maxwell) Chern-Simons
theories and brane configurations with $(p,q)$5-brane. For Coulomb
branches, the number of the massless adjoint scalar fields in various
supersymmetric theories exactly coincides with the number of the
freely moving directions of D3-branes stretched between two
5-branes. When we include a matter superfield into the supersymmetric
Chern-Simons theory two distinct symmetric and asymmetric phase
appear. The symmetric phase is peculiar to this Chern-Simons Higgs
system. We find the corresponding brane configuration for these
phases. We also identify the stringy counterpart of the topological
vortex state in the asymmetric phase.

\end{titlepage}
\newpage

\renewcommand{\thefootnote}{\arabic{footnote}}
\setcounter{footnote}{0}

\section{Introduction}

Recent developments in string theory are understandings of the
correspondence between the brane dynamics and the physics of gauge
theories in various dimensions. Hanany and Witten have first pointed
out \cite{HW} that $N{=}4$ supersymmetric gauge theory in
three-dimensions is realized as worldvolume effective theory on
D3-branes stretched between two NS5-branes and the ``mirror symmetry''
can be understand as a consequence of $SL(2,\Z)$ duality of Type IIB
superstring theory.

Their construction is extended to different numbers of supersymmetries 
\cite{dBHOY,dBHO} and gauge theory in various dimensions (for a review
see \cite{GK,Karch}). The advantage of this approach is that we can
geometrically analyze the structure of the moduli spaces of vacua as
the configuration space of branes. The geometrical construction of
moduli space make easy to classify possible supersymmetric vacuum.

From the point of view of brane solutions in Type IIB superstring
theory, the three-dimensional Hanany-Witten type configuration is
generalized to the case including rotated 5-branes at arbitrary angles
\cite{GGPT}. It contain the interesting 5-brane solution preserving $3/16$ of
supersymmetry. This means that in three-dimensions we have $N{=}3$
supersymmetric theory. One of the realization of the three-dimensional
$N{=}3$ supersymmetry is obtained by adding the Chern-Simons term
\cite{ZK,KL,Kao}. So the brane configuration with $3/16$ supersymmetry
can be associated with three-dimensional Chern-Simons theory.

Moreover, the systematic classification of possible Hanany-Witten type
configuration with various supersymmetries has been studied in
ref. \cite{KOO}. Since the Type IIB brane configuration with $3/16$
supersymmetry contains a $(p,q)$5-brane which is a bound state of NS5
and D5-brane, it is reasonable to consider that the replacement of one of the
NS5-branes with the $(p,q)$5-brane is related to adding the
Chern-Simons interaction. Under this identification, it is found that
the vacuum structure of Maxwell Chern-Simons theory is consistent with
the brane picture and a ``mirror symmetry'' is similarly realized by
Type IIB S-duality. This ``mirror symmetry'' in supersymmetric
Chern-Simons theory is recently
confirmed by using a generalized Fourier transformation in ref. \cite{KS}.

In this paper, we investigate the correspondence between the
$(p,q)$5-brane configuration introduced by ref. \cite{KOO} and
Yang-Mills (Maxwell) Chern-Simons theories in detail. More
specifically, we construct a explicit Lagrangian for the
supersymmetric Chern-Simons theories and identify vevs of emergent
fields with moduli parameters of Type IIB branes. This identification
gives us the knowledge to understand the complicated dynamics of
three-dimensional massive gauge theories and the dynamics of Type IIB
branes.

This paper is organized as follows: In section 2 we briefly review the
construction of the brane configurations with $(p.q)$5-brane. These
configurations in Type IIB theory are reduced from M5-brane
configurations in M-theory with various supersymmetries. We give all
possible Hanany-Witten type configuration with $N{\geq}2$ supersymmetry. 
In section 3, we see how to describe the Coulomb branch by vevs of
adjoint scalars in supersymmetric Chern-Simons theory. The number of
the adjoint scalar fields corresponds to the number of freely moving
directions of D3-branes. Finally, we consider the vacuum of Maxwell
Chern-Simons theory with matter superfields. In this model, there are
two distinct vacua. One of vacua, which is called as a symmetric
phase, exists on in the Maxwell Chern-Simons Higgs system. We can
find the brane configuration of the symmetric phase. We also identify
the topological vortex in the asymmetric phase as a bound state of
strings.

\section{Supersymmetric Configuration with $(p,q)$5-brane}

In this section we briefly review the construction of the brane
configuration with $(p,q)$5-brane.

The brane configurations in Type IIB string theory which are given by
Hanany and Witten \cite{HW} describe three dimensional supersymmetric
gauge theories. In this configuration D3-branes are suspended between
two parallel NS5-branes and the $3+1$ dimensional worldvolume
of D3-brane is compactified on a finite line segment. We denote the
extended directions of two NS5-branes and $N_c$ D3-branes as the
following symbols:
\beann
2\times {\rm NS5} & : & (012345),\\
N_c\times {\rm D3} & : & (012|6|),
\eeann
where the numbers in the parenthesis are the worldvolume directions
and the vertical lines on both side of $6$ mean that the worldvolume
of D3-branes are restricted to finite interval in the $x^6$-direction.
The ``left'' NS5-brane is located at $x^6=-L/2$ and the ``right'' is
at $x^6=L/2$.

The low-energy effective theory on $N_c$ parallel D3-branes is $3+1$
dimensional $SU(N_c)$ supersymmetric Yang-Mills theory in the limit that
string scale $l_s$ goes to zero, but this compactification on
the line reduces the D3-brane worldvolume theory to $2+1$
dimensions. Upon this Kaluza-Klein reduction on the line segment we
find the gauge coupling constant $g$ of this worldvolume theory is
given by
\be
\frac{1}{g^2}=\frac{L}{g_s},
\label{coupling}
\ee
where $g_s$ is a string coupling constant of Type IIB theory and $L$
is a length of the line segment, namely, a distance between two
NS5-branes in the $x^6$-direction.

In this configuration the presence of these two kinds of branes
preserves 8 of 32 supercharges of Type IIB string theory since each
kind of brane preserves half of supercharges. Therefore we have
$N{=}4$ supersymmetry in $2+1$ dimensions.

To add matter in the fundamental representation of the gauge group we
can introduce $N_f$ D5-branes, which have the worldvolume:
\beann
N_f\times{\rm D5} & : & (012789),
\eeann
without breaking any supersymmetries. The matter hypermultiplets come
from strings stretched between the $N_c$ suspended D3-branes and the
$N_f$ D5-branes. So $N_f$ represents the number of flavors.

In addition to these types of branes there also exists the
$(p,q)$5-brane which is a bound state of NS5- and D5-branes. One
of the extension of the above setup is replacement of the right
NS5-brane in the Hanany-Witten configuration with this $(p,q)$5-brane.
However, if we simply substitute the $(p,q)$5-brane for the right
NS5-brane and the left NS5-brane and the right $(p,q)$5-brane
is parallel then all supersymmetry is broken. To preserve some of the
supercharges we must rotate the $(p,q)$5-brane by some angles.

In order to count the residual supercharges we now lift the above Type
IIB configuration to M5- and M2-brane configurations in
M-theory. T-duality in the $x^2$-direction maps the NS5-brane and
the D5-brane into the NS5-brane and the D4-brane in Type IIA theory,
respectively. By lifting to M-theory these objects become
M5-branes. The NS5-brane corresponds to the M5-brane wrapping on the
$x^2$-direction and the D5-brane corresponds to the M5-brane wrapping
on the 11th-direction $x^{10}$. So two cycles of the torus $T^2$ with
coordinates $(x^2,x^{10})$ are related to NSNS-charge and RR-charge in
Type IIB theory. Similarly, $(p,q)$5-brane in Type IIB theory is also
unified as a M5-brane which is obliquely wrapping on the
torus $T^2$ at some angle $\theta$. This angle and charges of the
$(p,q)$5-brane are related each other as follows
\be
\tan\theta=\frac{2\pi R_{10}p}{2\pi R_{2}q}=g_s\frac{p}{q},
\label{tan}
\ee
where $R_{2}$ and $R_{10}$ are radii of the torus $T^2$. 

Using the above procedure D3-branes in Type IIB theory become
M2-branes stretched between two M5-branes in M-theory. In this situation
the right M5-brane tilted by the angle $\theta$ can be
rotated by more three angles in the configuration space of M-theory.
So, we have a general configuration in M-theory:
\beann
{\rm M5} & : & (012345)\\
N_c\times {\rm M2} & : & (01|6|)\\
{\rm M5}' & : & (01
\rot{2}{10}{\theta}
\rot{3}{7}{\psi}
\rot{4}{8}{\varphi}
\rot{5}{9}{\rho}
),
\eeann
where the symbol $\rot{\mu}{\nu}{\theta}$ means that the extended
direction of the ${\rm M5}'$-brane are rotated by the angle $\theta$ in the
$(x^\mu,x^\nu)$-space. If the angle $\theta$ is zero ${\rm M5}'$-brane become
a rotated NS5-brane in Type IIB configuration space. On the other
hand, the case of
$\theta=\pi/2$ corresponds to a D5-brane.

The presence of these branes imposes the constrains on the
11-dimensional Killing spinors $\epsilon$ \cite{OT}
\bea
{\rm M5} &:& \Gamma_{012345}\epsilon = \epsilon,
\label{M5}\\
{\rm M2} &:& \Gamma_{016}\epsilon = \epsilon,
\label{M2}\\
{\rm M5}'&:& R\Gamma_{012345}R^{-1}\epsilon = \epsilon,
\label{M5'}
\eea
where $R$ is the rotation matrix in the spinor representation and give 
by
\[
R=\exp\left\{
\frac{\theta}{2}\Gamma_{2,10}
+ \frac{\psi}{2}\Gamma_{37}
+ \frac{\varphi}{2}\Gamma_{48}
+ \frac{\rho}{2}\Gamma_{59}
\right\}.
\]
The numbers of the remaining supersymmetry are obtained by solving
eqs. (\ref{M5})-(\ref{M5'}) simultaneously. All possible solutions are 
completely classified in ref. \cite{KOO} by setting some relations
between four angles and we have $1/16$, $1/8$,
$3/16$ and $1/4$ of the original 32 supercharges. These fractions of
supersymmetry correspond to
$N{=}1,2,3,4$ supersymmetries, respectively, in $2+1$ dimensions.

In present paper we treat only the configuration for $N{\geq}2$
supersymmetry. This is because quantum corrections of higher
supersymmetric is suppressed rather than the $N{=}1$ case and it is
not so complicated to find the correspondence to the branelogy.

$N{\geq}2$ supersymmetric configurations are divided into the following
two cases. One is the NS5 configuration with $N{=}2,4$ supersymmetry
where the right 5-brane is the NS5-brane. The other is the
configuration including $(p,q)$5-brane with $N{=}2,3$.

We first consider the NS5 configuration. For $N{=}4$ configuration we
must set all angles to zero, $\theta=\psi=\varphi=\rho=0$. The right
${\rm M5}'$-brane is a NS5-brane in Type IIB theory, which is parallel to the
left NS5-brane. This is nothing but the Hanany-Witten configuration.
As we mentioned above, this configuration describes $N{=}4$
$SU(N_c)$ supersymmetric Yang-Mills theory. This theory includes three
real scalar fields in adjoint representation. Vevs of these adjoint
scalars and scalars dual to the gauge fields parametrize the Coulomb
branch. On the other hand, D3-branes in the brane configuration can
freely move along the $(x^3,x^4,x^5)$-space. These positions of
D3-branes on NS5-brane correspond to the vevs of real adjoint scalars. 
In addition the $SO(3)$ rotation group in the $(x^3,x^4,x^5)$ can be
identified with a part of the $R$-symmetry $SU(2)_V$ of $N{=}4$
supersymmetry algebra in three dimensions. The other part of the
$R$-symmetry $SU(2)_H$ corresponds to rotations in the
$(x^7,x^8,x^9)$-space.

The $N{=}2$ NS5 configuration is obtained by a rotation of the right
NS5-brane. In this rotation we must set same values on two pairs of
angles. Since $\theta$ is zero in NS5 configuration one of our choice
for angles is $\theta=\rho=0$ and $\varphi=\psi\neq0$. The special
case of $\varphi=\psi=\pi/2$ is investigated in ref. \cite{dBHOY} and
this brane configuration space well describes the moduli space of
vacua of $N{=}2$ supersymmetric Yang-Mills theory. The Coulomb branch
of this theory is parametrized by vevs of one real massless adjoint
scalar and dual gauge fields. In this configuration D3-branes can not
move in the $(x^3,x^4)$-space any longer since two NS5-branes are
completely twisted in that space. Then the vev of real adjoint scalar
corresponds to the positions of D3-branes in $x^5$-direction. For the
general value of $\psi$ this rotation angle is related to a mass of
two real adjoint scalars \cite{EGK,Barbon}, which originally belong to
the $N{=}4$ vector multiplet. Therefore we can identify the rotation
of NS5-brane with the supplement of the mass term of two adjoint
scalars into $N{=}4$ theory. Actually, this brane rotation breaks the
$SU(2)_V \times SU(2)_H$ rotation symmetry to $SO(2)\simeq U(1)_R$,
which coincide with the $N{=}2$ $R$-symmetry group.

Let us next mention about the configuration with $(p,q)$5-brane. This
configuration is realized by setting $\theta\neq0$.

A maximal supersymmetric configuration with $(p,q)$5-brane is the case
of $\theta=\psi=\varphi=-\rho$. This configuration preserves $N{=}3$
supersymmetry in $2+1$ dimensions. Since the all rotation angles are
same except for the signature, the rotation symmetry $SU(2)_V\times
SU(2)_H$ in the $(x^3,x^4,x^5)$- and $(x^7,x^8,x^9)$-spaces breaks to
a diagonal $SU(2)_D$. This rotational symmetry $SU(2)_D$ can be considered
as the $R$-symmetry of the three-dimensional $N{=}3$ supersymmetry algebra.

$N{=}2$ configuration is given by setting
$\rho=-\theta\neq\varphi=\psi$, that is, two different sets of same
angles. If we take $\theta=\rho=0$, then we have the $N{=}2$ NS5
configuration as mentioned above. Similarly, we can choose as
$\varphi=\psi=0$. In this case the right 5-brane is $(p,q)$5-brane but
still $N{=}2$ supersymmetric.

It is field theoretically shown that the three-dimensional $N{=}3$
maximal supersymmetric system can be constructed by adding
Chern-Simons terms \cite{ZK,KL,Kao}. So one expect \cite{GGPT} that
the worldvolume effective theory on this $2+1$-dimensional
intersection of rotated 5-branes is gauge theories with Chern-Simons
terms. In fact, since there exists a non-trivial vev of axion
(RR-scalar) field $C_0$ around $(p,q)$5-brane in Type IIB theory, the
coupling $\int C_0 F\wedge F$ on the D3-brane induces the Chern-Simons
term with coupling $\kappa=p/q$ \cite{KOO}.

In the following section, we investigate the correspondence between
these supersymmetric configurations with the $(p,q)$5-brane and the
moduli space of vacua of Chern-Simons theories.

\section{Coulomb Branches of The Non-Abelian Theories}

\subsection{$N{=}3$ supersymmetric Yang-Mills Chern-Simons theory}

We first begin with the $N{=}3$ configuration. The configuration in
Type IIB theory is
written as follows in the notation of the previous section:
\beann
{\rm NS5} & : & (012345)\\
N_c\times {\rm D3} & : & (012|6|)\\
(p,q)5 & : & (01
\rot{3}{7}{\theta}
\rot{4}{8}{\theta}
\rot{5}{9}{-\theta}
),
\eeann
where $\theta=\tan^{-1}\left(g_s\frac{p}{q}\right)$.  There originally
exists $SU(N_c)$ Yang-Mills gauge theory on the D3-branes. As we have
explained, the right $(p,q)$5-brane provides the Chern-Simons term
into the Yang-Mills theory. So we have the Yang-Mills Chern-Simons
theory as the effective theory on this configuration. The coupling
constant of Chern-Simons term is given by $\kappa=p/q$. However, the
coupling constant of the non-Abelian Chern-Simons theory should be a
integer in order to preserve gauge invariance under a large
transformation. Therefore we set the charges of $(p,q)$5-brane as
$(p,q)=(n,1)$.

$N{=}3$ supersymmetric field theory in three dimensions has the
$SU(2)_D$ $R$-symmetry. The vector multiplet contains one spin 1
massive vector
field $A_\mu$, three spin $1/2$ spinors $\lambda_i$, three spin 0 real
adjoint fields $X_i$ and one spin $-1/2$ spinor $\chi$. Three spinors and
adjoint scalars form a triplet under $SU(2)_D$.
The bosonic part of the Lagrangian of this $N{=}3$ supersymmetric
Yang-Mills Chern-Simons
theory is given by \cite{Kao}
\bea
{\cal L}_{\rm B}&=&
-\frac{1}{g^2}\Tr\left\{
\frac{1}{2}{F_{\mu\nu}}^2
+\left(\nabla_\mu X_i\right)^2
+\frac{1}{2}\left[X_i,X_j\right]^2
-\left(\frac{\kappa g^2}{4\pi}\right)^2
{X_i}^2
\right\}\\
&&+\frac{\kappa}{4\pi}\Tr\left\{
\epsilon^{\mu\nu\rho}\left(A_\mu\del_\nu A_\rho+\frac{2}{3}iA_\mu
A_\nu A_\rho\right)
-\frac{i}{3}\epsilon^{ijk}X_i\left[X_j,X_k\right]
\right\}.
\eea
Due to the Chern-Simons term the vector field topologically gains mass
$M=\left|\frac{\kappa g^2}{4\pi}\right|$ which is the same mass as the
adjoint scalars have.  Note that in the limit of $\kappa\rightarrow0$,
which corresponds to the limit $\theta\rightarrow0$ in the brane
configuration, one obtains the $N{=}4$ supersymmetric pure Yang-Mills
Lagrangian.

The Coulomb branch of supersymmetric gauge theory is described by
solutions of the following flatness condition for the adjoint scalar
fields
\be
M^2 X_i
-\frac{M}{2}
i\epsilon_{ijk}\left[X_j,X_k\right]
-\left[X_j,\left[X_i,X_j\right]\right]=0.
\label{flat}
\ee
For real $X_i$, the general solution of this condition is
$\left<X_i\right>=0$. So the Coulomb branch of this theory is
completely lifted because of the mass term coming from the
Chern-Simons interaction.

From the point of view of the brane configuration, it is easy to
understand this Coulomb branch moduli as the following. In the $N{=}4$
brane configuration, $N_c$ D3-branes can freely moved on the
$(x^3,x^4,x^5)$-space since two NS5-branes are parallel. So we can
identify positions of D3-branes in the $(x^3,x^4,x^5)$-space with vevs
of the three real adjoint scalar fields, which describe a part of the
Coulomb branch.

However, in the $N{=}3$ configuration, since the left NS5-brane and
the right $(p,q)$5-brane is completely twisted in $(x^3,x^4,x^5)$- and
$(x^7,x^8,x^9)$-space, D3-branes can not move anywhere. This means
that the corresponding vevs of the adjoint scalars must be at origin,
namely, $\left<X_i\right>=0$, which coincide with the solution of the
flatness condition (\ref{flat}).

The mass of the adjoint scalars can be written in terms of the rotation
angle of the $(p,q)$5-brane as
\be
\left|\frac{\kappa g^2}{4\pi}\right|=\frac{1}{4\pi L}|\tan\theta|,
\ee
where we use the relation (\ref{coupling}) and (\ref{tan}). This
relation also agree with the results in ref. \cite{EGK,Barbon}.

In the $\theta\rightarrow 0$ limit the mass of the adjoint scalar
fields vanishes and at the same time the Chern-Simons interaction is
dropped. In the brane configuration the right $(p,q)$5-brane become
NS5-brane and to be parallel to the left NS5. So we have $N{=}4$
supersymmetric pure Yang-Mills theory. On the other hand, in the limit
$\theta\rightarrow \pi/2$ the mass of the vector multiplet goes to
infinity and all gauge degrees of freedom are
decoupled. In this limit the right $(p,q)$5-brane become D5-brane which 
extends in the $(x^0,x^1,x^2,x^7,x^8,x^9)$-space. This configuration
also recover the $N{=}4$ supersymmetry again and appears in the Higgs
branch of the Hanany-Witten configuration \cite{HW}.

\subsection{Breaking to $N{=}2$}

We next consider the supersymmetry breaking of $N{=}3$ theory down to $N{=}2$.
In the brane language this is done by making two of the angles
different from the others. Namely, we have the configuration
\bea
{\rm NS5} & : & (012345)\nn\\
N_c\times {\rm D3} & : & (012|6|)\\
(p,q)5 & : & \left(012
\rot{3}{7}{\psi}
\rot{4}{8}{\psi}
\rot{5}{9}{-\theta}
\right).\nn
\eea
As the result, the corresponding mass of the adjoint scalars become different.

We can field theoretically understand this supersymmetry
breaking. Originally $N{=}3$ vector multiplet contains the fields
$(A_\mu,\lambda_i,X_i,\chi)$. If we give the different mass $\mu$ to
the $X_1,X_2$ and $\lambda_3,\chi$, these fields form the $N{=}2$
neutral chiral supermultiplet $\Phi$. The rests of the fields
$(A_\mu,\lambda_a,X_3)$ are
contents of the $N{=}2$ vector multiplet, which still have the mass
$\left|\frac{\kappa g^2}{4\pi}\right|$.

In this $N{=}2$ theory three adjoint scalars are massive as long as
$\psi\neq0$. So the Coulomb branch of this theory is still lifted and
there is no moduli. This corresponds to the twisting configuration of
the NS5-brane and the rotated $(p,q)$5-brane. However, if we set
$\psi=0$ the NS5-brane and $(p,q)$5-brane become parallel in the
$(x^3,x^4)$-space. Therefore D3-branes can now freely move in the
$(x^3,x^4)$-space. In this configuration, two adjoint scalars $X_1$
and $X_2$ become massless and the Coulomb branch appears.

\section{Broken and Unbroken Phases of Abelian Theory}

In this section we consider the Higgs branches of the $N{=}2$
supersymmetric Chern-Simons theory. We first briefly review the
description of the Higgs branch in the Hanany-Witten
configuration. The inclusion of the matter hypermultiplets corresponds
to the addition of the D5-branes, which extend in the
$(x^0,x^1,x^2,x^7,x^8,x^9)$-space, between two NS5-branes. The matter
hypermultiplets come from the open string stretched between D5-branes
and D3-branes\footnote{There is a couple of chiral fields with
opposite chirality with respect to orientation of strings.}. For
simplicity, we consider only Abelian theory with one matter
hypermultiplet, which includes two complex scalar fields $(q,\til{q})$ from now on.

If we denote the distance between the D5-brane and the D3-brane in the 
$(x^3,x^4,x^5)$-space as $d$, the mass of the hypermultiplet is given
by
\be
m = T_{\rm F1}d
 =  \frac{d}{2\pi {l_s}^2},
\ee
where $T_{\rm F1}$ is the string tension and $l_s$ is the string
length. In the decoupling limit of massive excitation of strings
$l_s\rightarrow 0$, the quantity of $m$ must be fixed.

The Higgs branch of this theory emanates from the point where the
hypermultiplet is massless, namely, $d=0$ in the brane
configuration. When $d=0$, the D5-brane can separate the
D3-brane into two parts. As the result, two NS5-branes can shift each
other's positions in the $(x^7,x^8,x^9)$-space. We denote this
difference as $\til{d}$.

The gauge symmetry in this branch is completely broken since the
D3-brane is divided into two parts by the D5-brane and all gauge
degrees of freedom are decoupled. In the broken phase of the Abelian
gauge theory, there exists a vortex state, which is an
electric-magnetic dual object of the fundamental matter. Since the
electric-magnetic duality of the brane effective theory is achieved by
the Type IIB S-duality, the stringy counterpart of the vortex state is
a D-string stretched between two separated parts of D3-brane. The mass
of this state is
\be
\til{m} = T_{\rm D1}\til{d}
=\frac{\til{d}}{2\pi {\til{l_s}}^2},
\ee
where $\til{l_s}={g_s}^{1/2} l_s$ is a string length of the dual
theory. Filed theoretically, the mass of the vortex state is given by
the Fayet-Iliopoulos (FI) parameter $\zeta$. If we define as
\be
\zeta=\frac{\til{m}}{4\pi}=\frac{\til{d}}{8\pi^2{\til{l_s}}^2},
\label{vortex mass}
\ee
the FI
parameter is proportional to the difference of two NS5-brane
positions in the $x^9$-direction.

Next, we consider a similar situation in the $N{=}2$ brane
configuration with $(p,q)$5-brane. We add only one set of massless
matter superfields which plays a role of Higgs field and consider the
Abelian case, that is, the Maxwell
Chern-Simons Higgs system. Moreover, we set the angles $\psi$ to
$\pi/2$ in the $N{=}2$ brane configuration. So the brane configuration is
as follows
\beann
{\rm NS5} & : & (012345)\\
{\rm D3} & : & (012|6|)\\
{\rm D5} & : & (012789)\\
(p,q)5 & : & \left(01278\rot{5}{9}{-\theta}\right).
\eeann
The two adjoint scalars
$X_1$ and $X_2$ are decoupled from the theory and so we can easily pay
attention to the true nature of the Higgs phase of the Abelian theory.

The squark Higgs fields $q,\til{q}$ in the matter superfields
couple to a non-dynamical D-field in the off-shell vector
multiplet. If we integrate out the auxiliary D-field we have a bosonic
scalar potential
\be
V=-\frac{g^2}{2}\left(
|q|^2-|\til{q}|^2-\zeta+\frac{\kappa}{4\pi} X_3
\right)^2
-{X_3}^2\left(|q|^2+|\til{q}|^2\right),
\label{potential}
\ee
where $\zeta$ is the FI parameter.

The potential of the aforementioned Maxwell Higgs theory is obtained
by setting $\kappa=X_3=0$. In this case there is only the asymmetric
phase $|q|^2-|\til{q}|^2=\zeta$. However, the potential
(\ref{potential}) admits two distinct vacua, an asymmetric phase and a
symmetric phase. We examine the correspondence between each branch and
the brane configuration from now on.\\\\
\underline{\it asymmetric phase}

\begin{figure}
\centerline{
\includegraphics[scale=0.9]{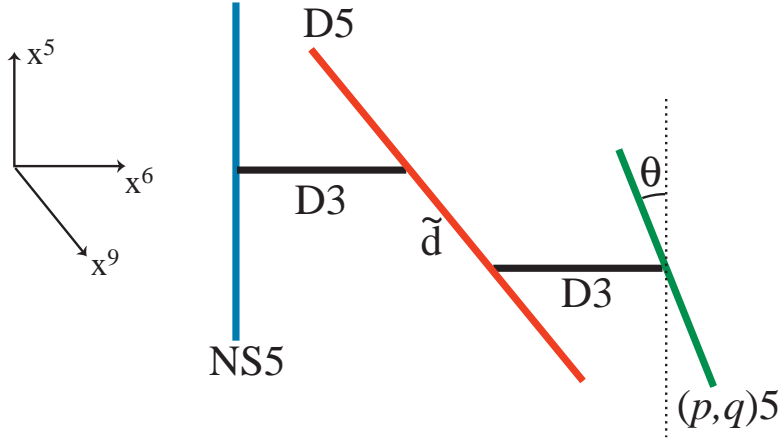}
}
\caption{The brane configuration in the asymmetric phase.}
\label{asymmetric}
\end{figure}

The asymmetric phase is described by $|q|^2-|\til{q}|^2=\zeta$ and
$X_3=0$. This phase is the same as the asymmetric phase in the
previous case. So the corresponding configuration is also very
similar to the Maxwell Higgs one. Namely, the D5-brane divide the
D3-brane into two segments. One of the segments of the D3-brane
stretch between the left NS5-brane and the D5-brane and the other is
between the D5-brane and the right $(p,q)$5-brane. In this phase the
relative position of the 5-branes along the $x^9$-direction can differ 
because of this split of the D3-brane and the gauge
symmetry on the D3-brane is broken simultaneously. We denote the
difference of 5-branes in the $x^9$-direction as $\til{d}$.
(See Fig.\ref{asymmetric}.)

There also exists the topological vortex state in this broken phase of the
Maxwell Chern-Simons Higgs system. However, in the $(p,q)$5-brane
configuration we can not simply identify this vortex state with the
D-string stretch between two fragments of the D3-brane. From the SUGRA 
solution of the $(p,q)$5-brane, there is a non-trivial vev of the
axion field like as \cite{KOO}
\be
\chi(x^6,x^7,x^8,x^9)=
\frac{1}{g_s}
\frac{\sin\theta\cos\theta(H^{1/3}-H^{2/3})}
     {H^{-1/3}\sin^2\theta+H^{2/3}\cos^2\theta},
\ee
where $H=1+\frac{{l_s}^2}{r^2}$ is a harmonic function on the
$(x^6,x^7,x^8,x^9)$-space and $r$ is a
distance from the $(p,q)$5-brane. We now assume that the distance $L$
between two 5-branes is satisfied $L\ll l_s$\footnote{This limit is
connected with the pure Chern-Simons limit ($\kappa$ is fixed and
$g\rightarrow\infty$).}. Since the D5-brane sits between two 5-branes,
the distance from $(p,q)$5-brane $r$ is also satisfied $r\ll l_s$. In
this limit, that is, the axion field take a
constant value near the D5-brane
\be
\chi=-\frac{1}{g_s}\tan\theta=-\kappa.
\ee
In this background, D-string must appear as a dyon with electric
charge proportional to $\chi$ by the Witten effect. In fact,
since the tension of $(a,b)$-string in this background is given by
\cite{Schwarz}
\be
T_{(a,b)}
=\frac{1}{2\pi{l_s}^2}
\sqrt{\left(\frac{b}{g_s}\right)^2+\left(a+b\chi\right)^2},
\ee
the D-string, {\it i.e.} $(0,1)$-string, is not lightest
state. A lighter state is a $(\kappa,1)$-string with the tension
\be
T_{(\kappa,1)}=\frac{1}{2\pi{l_s}^2 g_s},
\ee
which is the same tension as the D-string in the Maxwell theory
configuration. Therefore the vortex state near the $(p,q)$5-brane
configuration is not only magnetically charged but also electrically
charged.

Field theoretically, the Chern-Simons interaction gives the relation
between electric charge $Q$ and magnetic flux $\Phi$
\be
Q=\kappa \Phi
\ee
from the Gauss law constraint. So the vortex solution carry both
magnetic charge and electric charge in the ratio of 1 to
$\kappa$. This consequence exactly agree with the $(\kappa,1)$-string
state around the $(p,q)$5-brane background.\\\\
\underline{\it symmetric phase}

\begin{figure}
\centerline{
\includegraphics[scale=0.8]{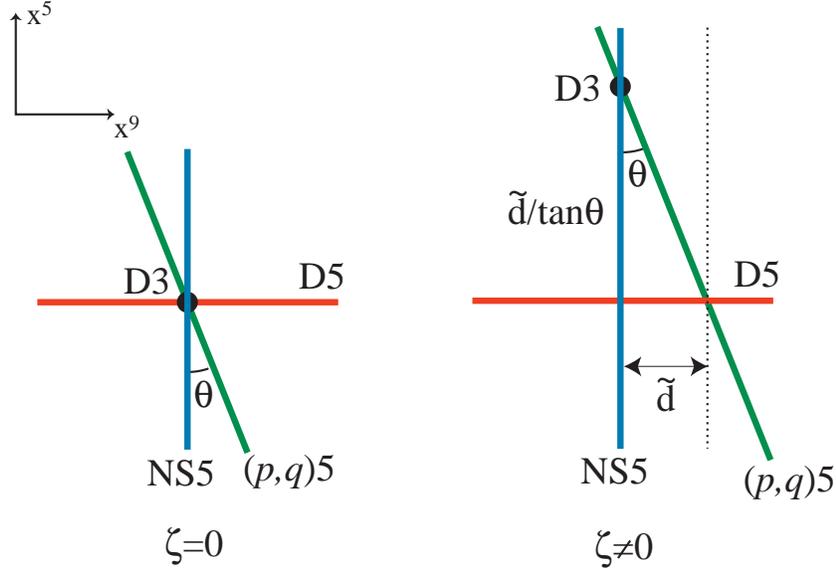}
}
\caption{The brane configuration in the symmetric phase}
\label{symmetric}
\end{figure}

For the symmetric phase the vevs of the scaler fields are
\bea
&&q=\til{q}=0,\\
&&X_3=4\pi\zeta/\kappa.
\label{sym vev}
\eea
This phase does not exist in the Maxwell Higgs system. And also there
is no corresponding configuration in the $N{=}4$ Hanany-Witten
setup. However, in our case there exists another possible
vacuum configuration for $\zeta\neq0$ due to the rotation of the
5-brane. If we want to set $\zeta\neq0$ we must shift the positions of
the 5-branes each other along the $x^9$-direction. In general, the
D3-brane suspended between the 5-branes prevents this motion. One of
settlements is the division of the D3-brane into two parts as we
mentioned before. Another solution is that the D3-brane slide up
along the $x^5$-direction and is attached again on the intersecting
point of the NS5-brane and the rotated $(p,q)$5-brane in the
$(x^5,x^9)$-plane. (See Fig. \ref{symmetric}.)

In this symmetric phase configuration, the position of the D3-brane in
the $x^5$ coordinate is $x^5=\til{d}/\tan\theta$ where $\til{d}$
is a gap of the 5-brane positions. The gap $\til{d}$ corresponds to
the FI parameter $\zeta$ as in eq. (\ref{vortex mass}) and the $x^5$
position of the D3-brane corresponds to the vev of the adjoint scalar
$X_3$. So we find that the adjoint scalar vev is given by
\bea
X_3&=&\frac{\til{d}}{2\pi {l_s}^2\tan\theta}\\
   &=&\frac{4\pi\zeta}{p/q}.
\eea
This result is exactly agree with (\ref{sym vev}).

Moreover, the distance between the D3 and D5-brane is
$\til{d}/\tan\theta$. This means that the mass of the matter
superfields which is coming from the fundamental string between D3 and
D5-brane is $4\pi\zeta/\kappa$. Correspondingly, when the $X_3$ field has
the vev (\ref{sym vev}) the scalar potential (\ref{potential}) gives
the mass $4\pi\zeta/\kappa$ to the $q$ and $\til{q}$. This is also
consistent with the brane picture.

Finally, we would like to comment on vortex states in this phase. In
this symmetric phase, it is known that there exists a non-topological
vortex \cite{JW,JLW}. However, since charges of the non-topological vortex
contain a continuous parameter, it seems to be difficult to find the
corresponding state in string theory.

\section*{Acknowledgments}

I would like to thank T. Kitao and N. Ohta for arguments during the
early stage of this work. I also grateful to B.-H. Lee for useful
discussions and comments at Niseko Winter School.

\input refs.tex

\end{document}

%% file: refs.tex
%
%

\newcommand{\NP}[3]{Nucl.\ Phys.\ {\bf #1} (#2) #3}
\newcommand{\PL}[3]{Phys.\ Lett.\ {\bf #1} (#2) #3}
\newcommand{\PR}[3]{Phys.\ Rev.\ {\bf #1} (#2) #3}
\newcommand{\PRL}[3]{Phys.\ Rev.\ Lett.\ {\bf #1} (#2) #3}
\newcommand{\hepth}[1]{{\tt hep-th/#1}}

\newcommand{\lit}[3]{#1, ``#2,'' #3}